\documentclass[referee]{raa_rb}
\usepackage{graphicx,times}
\usepackage{natbib}
\usepackage{amssymb,amsmath}
\bibpunct{(}{)}{;}{a}{}{,}
\usepackage{verbatim}

\usepackage[a4paper=true,pagebackref=true]{hyperref}
\hypersetup{pdftitle = The title of my PDF, pdfauthor = My name, pdfsubject= The subject, pdfkeywords = keyword1 keyword2 keyword3}
\hypersetup{colorlinks = true, linkcolor = green, anchorcolor = red, citecolor = blue, filecolor = red, pagecolor = red, urlcolor = red}

\begin{document}
	
	\title{Binary fraction of O and B-type stars from LAMOST data
	}
	
	\volnopage{Vol.0 (20xx) No.0, 000--000}      
	\setcounter{page}{1}          
	
	\author{Feng Luo
		\inst{1,2}
		\and Yong-Heng Zhao
		\inst{1,2}
		\and Jiao Li
		\inst{1,2}
		\and Yan-Jun Guo
		\inst{2,4}
		\and Chao Liu
		\inst{2,3}
	}
	
	\institute{National Astronomical Observatories, Chinese Academy of Sciences,
		Beijing 100012, China; 
		\and
		University of Chinese Academy of Sciences, Beijing 100049,  China;
		\and
		Key lab of Space Astronomy and Technology, National Astronomical Observatories, Beijing, 100101, China;
		{\it liuchao@nao.cas.cn}
		\and
		Yunnan observatories, Chinese Academy of Sciences, P.O. Box 110, Kunming, 650011, China; 
		\vs\no
		{\small Received~~20xx month day; accepted~~20xx~~month day}
	}
	
	\abstract{
		Binary stars plays important role in the evolution of stellar populations .
		The intrinsic binary fraction ($f_{bin}$) of O and B-type (OB) stars in LAMOST DR5 was investigated in this work.
		We employed a cross-correlation approach to estimate relative radial velocities for each of the stellar spectra. 
		The algorithm described by \cite{2013A&A...550A.107S} was implemented and several simulations were made to assess the performance of the approach.
		Binary fraction of the OB stars are estimated through comparing the uni-distribution between observations and simulations with the Kolmogorov-Smirnov tests.
		Simulations show that it is reliable for stars most of whom have $6,7$ and $8$ repeated observations.
		The uncertainty of orbital parameters of binarity become larger when observational frequencies decrease.
		By adopting the fixed power exponents of $\pi=-0.45$ and $\kappa=-1$ for period and mass ratio distributions, respectively, we obtain that $f_{bin}=0.4_{-0.06}^{+0.05}$ for the samples with more than 3 observations. 
		When we consider the full samples with at least 2 observations, the binary fraction turns out to be $0.37_{-0.03}^{+0.03}$. 
		These two results are consistent with each other in $1\sigma$.
	}
	
	\maketitle

	%
	\section{Introduction}           
	\label{sect:intro}
	
	Binary stars play a crucial role in the evolution of stars and galaxies(\citealt{2012MNRAS.424.1925C}, \citealt{2017A&A...598A..84A}). 
	Almost a half of solar-type stars locate in binary systems
	(\citealt{2010ApJS..190....1R},\citealt{2017ApJS..230...15M}).
	A nearby companion would probably affect the evolution of massive stars in binary systems (\citealt{1992ApJ...391..246P}, \citealt{2008IAUS..250..167L}, \citealt{2011MNRAS.414.3501E}), leading to phenomena such as stellar mergers, X-ray binaries and gamma-ray bursts (\citealt{2012Sci...337..444S}).
	Therefore, it is non-trivial to identify binary system from single stars and to determine the orbital parameters of the binary stars in different Galactic environments (\citealt{2017MNRAS.469L..68G}).
	
	Compared the distance to earth, the separations of most binaries are too close to resolve in  photometry. 
	When orbital periods are relative short, 1000 days for example, there are two cases that binaries can be detected through spectroscopic approaches. 
	On one hand, spectra will appeared split or contained "double lines" if the primary and secondary have near degree of luminosities (\citealt{2004A&A...424..727P}, \citealt{2017PASP..129h4201F}, \citealt{2017A&A...608A..95M}). 
	Bimodal peaks would emerge in the cross-correlation function of spectra for these binary systems.
	On the other hand, the luminosity of secondary is much smaller than primary and it can not contribute enough flux to the spectrum. 
	But, it can lead to orbital motion of primary and bring distinguishable radial velocity (RV) variation (\citealt{2013ApJ...779..116M}, \citealt{2016AJ....151...85T}, \citealt{2017ApJ...837...20P}, \citealt{2018ApJ...854..147B}).
	In this paper, we studied the latter case.
    
	A few previous works have studied the intrinsic binary fraction $f_{bin}$ in different environments in the past decades.
	\cite{2010ApJS..190....1R} reported that $f_{bin}$ of solar-type stars is about $0.34$.
	OB stars are considered having a larger binary fraction in some literatures.
	For B-type stars, \cite{2012MNRAS.424.1925C} showed that the $f_{bin}$ is $0.46\pm0.03$.  \cite{2015AA...580A..93D} found that there is a relatively high $f_{bin}$ which is $0.58\pm0.11$.  
	For O-type stars, $0.42\pm0.04$ was came up with \cite{2015AJ....149...26A}, $0.50\pm0.03$ with \cite{2014ApJS..211...10S}, $0.51\pm0.07$ with \cite{2014ApJS..213...34K}, $0.68\pm0.03$ with \cite{2012MNRAS.424.1925C}, and $0.69\pm0.09$ with \cite{2012Sci...337..444S} respectively.
	The different results of $f_{bin}$ may caused by different methods or data samples adopted in previous researches.
	These studies seem to reveal that most early type stars are residing in binary systems.
	The results of binary fraction studied by above-mentioned works are listed in Table.\ref{table:Binary overview}.
	
	\begin{table}[h!]
		\centering
         \begin{tabular}{||c|c|c|c||} 
         \hline
         $f_{bin}$ & Number of samples & Spectral type & reference \\ [0.5ex] 
         \hline\hline
         $0.34\pm0.02$ & 454 & FGK & \cite{2010ApJS..190....1R} \\
         \hline
         $0.46\pm0.03$ & 226 & B & \cite{2012MNRAS.424.1925C} \\
         \hline
         $0.58\pm0.11$ & 408 & B & \cite{2015AA...580A..93D} \\
         \hline
         $0.42\pm0.04$ & 161 & O & \cite{2015AJ....149...26A} \\
         \hline
         $0.50\pm0.03$ & 194 & O & \cite{2014ApJS..211...10S} \\
         \hline
         $0.51\pm0.07$ & 45 & O & \cite{2014ApJS..213...34K} \\
         \hline
         $0.68\pm0.03$ & 243 & O & \cite{2012MNRAS.424.1925C} \\
         \hline
         $0.69\pm0.09$ & 71 & O & \cite{2012Sci...337..444S} \\ [1ex]
         \hline
         \end{tabular}
	\caption{Overview of recent studies of binary fraction of different samples.}
	\label{table:Binary overview}
    \end{table}
	LAMOST telescope can obtain at most 4000 spectra within one observation and hence can survey the sky in a very high efficiency (\citealt{2012RAA....12.1197C},  \citealt{2012RAA....12..723Z}).
    The LAMOST targets mainly on stars and its data is widely used in researches of stellar
    physics and the Milky Way. 
    Its data release 5 (DR5) contains 9,027,634 spectra obtained between October 24, 2011 and June 16, 2017.
	
	The primary goal of this study is to estimate the binary fraction of OB stars observed with LAMOST.
	At the same time, the distribution law of orbital parameters were explored.
	Observations of different epochs for a same source may obtain different radial velocities, then binaries can be discriminated from single stars from RV variation.
	We employ the method proposed by \cite[][hereafter S13]{2013A&A...550A.107S}, to estimate the $f_{bin}$ for the LAMOST OB stars with multiple observations.
	LAMOST sample and data selection are described in Section 2.
	Details of our method and validation are presented in Section 3.
	The processing results of observational data are analyzed in Section 4.
	Discussion and conclusion are given in Section 5 and 6, respectively.
	
	\section{Data}    \label{sec:Data}
	\subsection{Samples from LAMOST}    \label{subsec:Samples from LAMOST}
	\begin{figure}[htbp]
		\centering
		\includegraphics[width=1\linewidth,height=0.20\textheight]{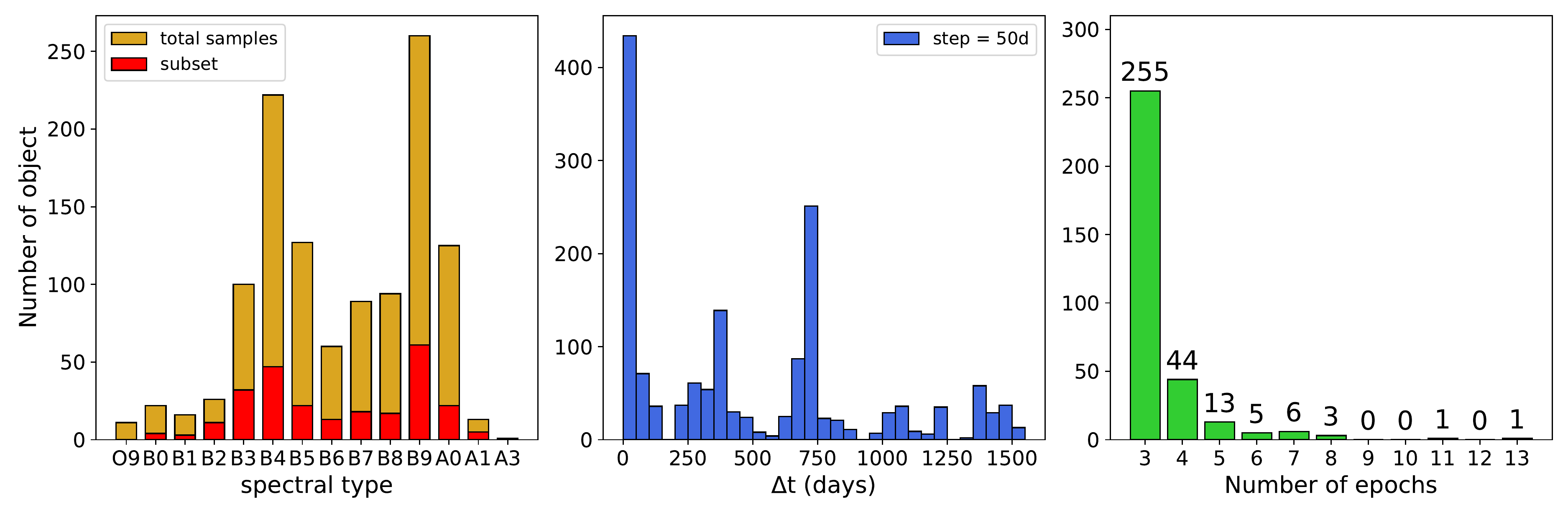}
		\caption{Left panel: 
		Distributions of the spectral type of the OB stars from \cite{2019ApJS..241...32L} and the subset for this work are presented with yellow and red colours, respectively.
		Middle panel: It displays the distribution of observational interval of the selected sub-samples. Right panel: It illustrates the distribution of the number of observation for the selected sub-samples.
		}
		\label{fig:basicInfo_fig1.pdf}
	\end{figure}
	A catalogue of OB stars from LAMOST released in 2019 (\citealt{2019ApJS..241...32L}) contains 22,901 spectra of 16,032 stars.
	1,715 of them have been targeted with 2 or more epochs by LAMOST.
	We discard several spectra which SNRg $< 10$ or have bad pixels, and leave 1,680 sources for this work.
	Where SNRg is the signal-to-noise ratio at g-band, which can be got from the header recorded in the corresponding fits file of each spectrum.
	Each spectrum was checked by eyes, and pixels with flux equals 0 were regarded as bad pixels.
	
	In general, for a given binary candidate, the more observations we have, the more orbital information we can obtain. 
	Several simulated experiments with mock data were made to assess the capability of repeated observation times to predict orbital parameters. 
	We generated 9 sets of mock samples, each set has the same sample size of 328, the same prior distributions of orbital parameters, but different observation epochs in each set of samples from $2 - 10$. 
	The simulated results show that the uncertainty of predictions of orbital parameters becomes smaller when the observational times increase.
	
	According to our experiments, samples who have observed only twice help little to predict the distribution of orbital parameters.	
	Therefore, 328 sources containing 1126 spectra with at least 3 repeated observations were chosen for this study.
	The distribution of spectral type for the OB stars given by the catalog is shown in the left panel of Fig. \ref{fig:basicInfo_fig1.pdf}. 
	Yellow bars indicate the whole samples, while red bars the subset to be used in binarity study.
    The distribution of time intervals for the repeated observations is shown in the middle panel of Fig.1.
	The right panel of Fig.\ref{fig:basicInfo_fig1.pdf} shows distribution of the number of observations of our samples, it shows that about $77.8\%$ stars have 3 repeated observations.
	\subsection{Relative radial velocity} \label{subsec:Relative radial velocity}
	LAMOST pipeline (\citealt{2015RAA....15.1095L})	provides the doppler shift of each spectrum, which could be used to estimate RV simply by multiplying light speed.
	However, the doppler shifts in LAMOST RVs are not sufficiently precise for early type stars, since their spectra have fewer lines.
	
	In order to get more precise RVs, we employed a cross-correlation approach to estimate the relative RVs.
	In general, the usage of template-matching method within a group of spectra corresponding to an observational source that observed at different time can avert the risk of inconsistency of spectral type between the spectrum to be measured and empirical template spectral library, as well as the error brought by this procedure.

	The wavelength range used in the cross-correlation analysis depends on the features of OB spectra being considered within the spectral region.
    Wavelength range of $3900 - 5000$ \AA~ was chosen because: 1) There are absorption lines such as H$_\beta$, H$_\gamma$ and H$_\delta$ in this range 
    which can be used to estimate the relative RV. 2) Many OB stars are brighter at this wavelength range.
    The wavelength calibration is by use of arc lines, a vacuum wavelength scale is applied for it (\citealt{2015RAA....15.1095L}).
    
\begin{figure}
		\centering
		\includegraphics[width=0.60\linewidth, height=0.30\textheight]{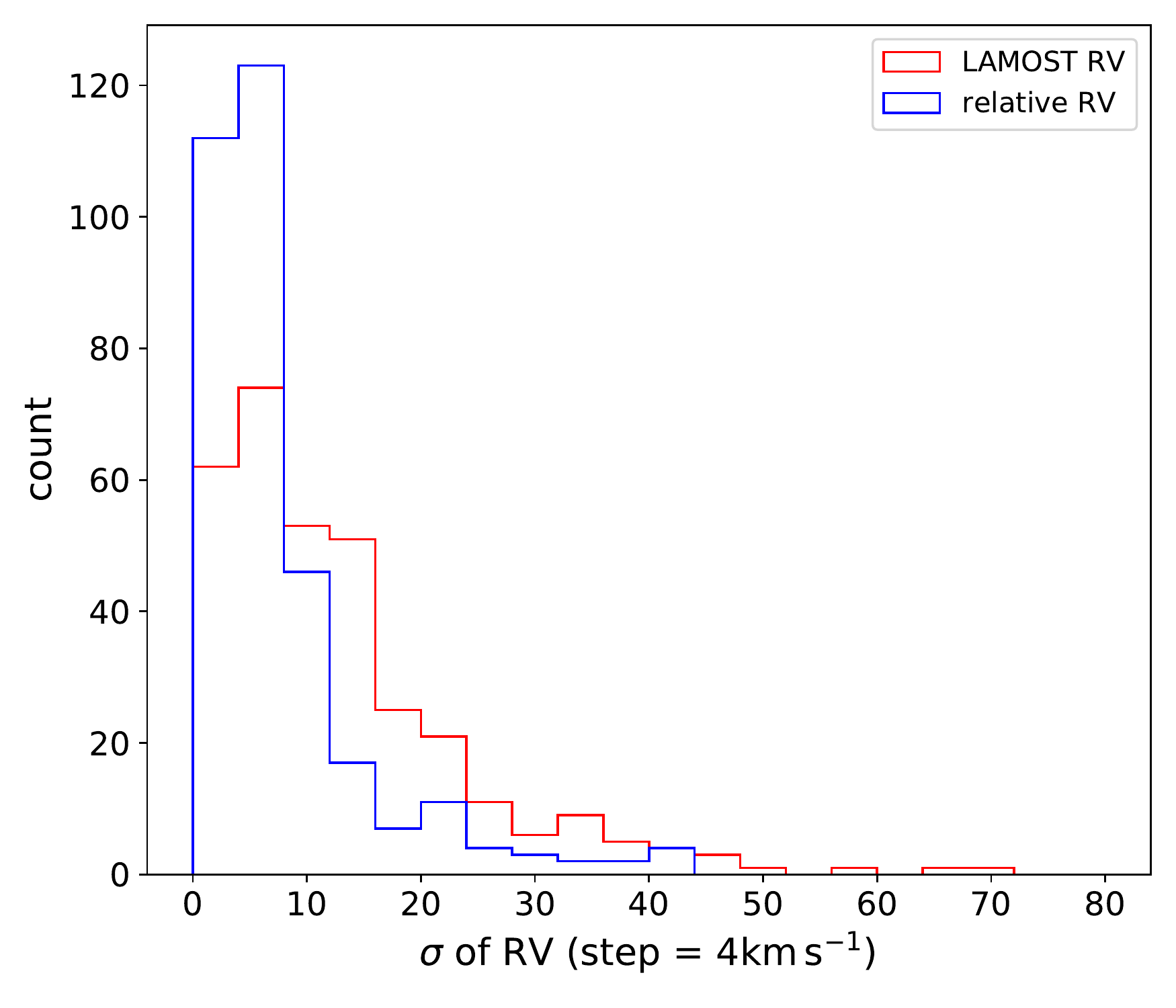}
		\caption{The uncertainty of RV measurement by cross-correlate method in this paper is about $0.7$ km\,s$^{-1}$. The blue line displays the distribution of $\sigma_{RV}$ measured from cross-correlation. The red line shows the distribution of $\sigma_{RV}$ provided by LAMOST catalog.}
		\label{fig:std_comparison_fig2.pdf}
	\end{figure}

	We chose the spectra with highest SNRg for each star as the template which relative radial velocity is fixed at zero.
	Then the rest spectra of the same star were respectively cross-correlated with the template to calculate relative RV.
	We defined $\sigma_{RV}$ as the standard deviation of RVs for a given source, which represents the dispersion of the RVs. 
	Results of $\sigma_{RV}$ calculated by this strategy is showed in Fig. \ref{fig:std_comparison_fig2.pdf}, blue line.
	As a contrast, the red line in Fig. \ref{fig:std_comparison_fig2.pdf} indicated the $\sigma_{RV}$ provided by LAMOST catalog.
	
	\subsection{Error of RV measurement} \label{subsec:Error of RV measurement}
	\begin{figure}
		\centering
		\includegraphics[width=0.85\linewidth, height=0.32\textheight]{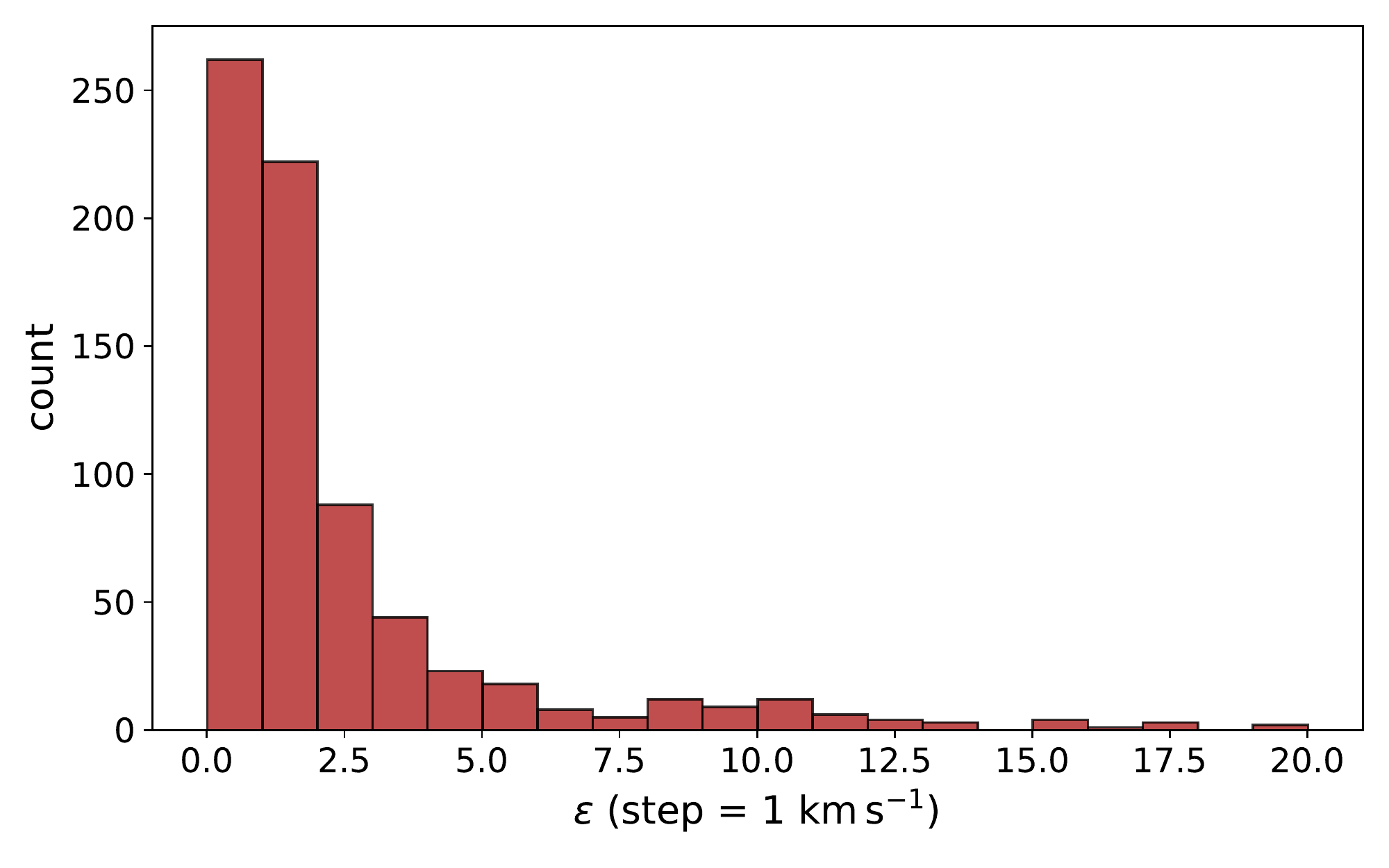}%
		\caption{Distribution of measurement error of RV. The template spectra in each group were not included.}
		\label{fig:mmes_328_fig3.pdf}
	\end{figure}
	Binaries can be detected through RV variations.
	However,the observed RV differences between epochs may occur only due to the measurement error.
	Hence, the results of radial velocity measurements for the same source are composed of both measurement error and doppler shift caused by orbital motion within binary systems.
	The determination of the measurement error of RV is particularly important because it plays a significant role in the accuracy of our final analysis results.
	
	The measurement error of RV as $\varepsilon$ for each spectrum was determined with a Monte-Carlo simulation.
	For each spectrum we draw 200 spectra with randomly added Gaussian noise.
	The arbitrarily drawn noise follows sigma as the following.
	\begin{equation}  \label{Eq:sigma_noise}
		\sigma_{\text {noise}}=\sqrt{\left(\frac{f_{tp}}{snrg_{la}}\right)^{2}-\left(\frac{f_{tp}}{snrg_{tp}}\right)^{2}}  ,
	\end{equation}
	where $f_{tp}$ is the flux of the template, $snrg_{la}$ and $snrg_{tp}$ are SNRg of spectra to be measured and the template, respectively.
	Then, each spectrum with random noise drawn by Eq.\ref{Eq:sigma_noise} was cross-correlated with its corresponding template to derive the relative RVs.
	Finally, the uncertainty of RV of each spectrum can be derived from the standard deviation of the RVs of the $200$ simulations.

	The $\varepsilon$ of each template was defined as $10^{-4}$ km\,s$^{-1}$ for the following two reasons.
	Firstly, to avoid the exception with a divisor of zero, and make sure that the calculations go on wheels.
	Secondly, it should be defined as a sufficiently small value. 
	The $\varepsilon$ of each spectrum was derived from the template-matching method with its corresponding template, but it will be very small for the template spectrum itself.
	We tested the algorithm by adopting $\varepsilon$ as $10^{-3}$ and $10^{-2}$ km\,s$^{-1}$, there is no change in the final results.
	The distribution of $\varepsilon$ of all sources but the templates is shown in Fig. \ref{fig:mmes_328_fig3.pdf}. 
	
	\section{Method} \label{sec:Method}

	\subsection{Introduction to S13 method}	\label{subsec:Introduction to Sana's method}

	\begin{table}[h!]
		\centering
		\begin{tabular}{||c | l||} 
			\hline
			Name & Comment\\ [0.5ex] 
			\hline\hline
			$P$ & orbital period  \\ 
			$q$ & mass ratio \\
			$e$ & eccentricity \\
			$m_{1}$ & mass of the primary \\
			$i$ & angle of inclination \\ 
			$\omega$ & longitude of the periastron \\ 
			$T_{0}$ & the time of periastron passage \\ 
			[1ex] 
			\hline
		\end{tabular}
		\caption{Kinematics orbital parameters of binary}
		\label{table:orbital parameters}
	\end{table}
	
	There are seven parameters that describe binary orbits as shown in Table.\ref{table:orbital parameters}
	The RV equation (Eq.\ref{Eq:RV}) can be drawn through the geometric calculation of the motion of elliptic orbit.
	\begin{equation} \label{Eq:RV}
		\mathrm{RV}=\frac{2 \pi a \sin i}{P \sqrt{\left(1-e^{2}\right)}}[\cos (\theta+\omega)+e \cos \omega]  + \, \gamma\,\add{,} 
	\end{equation}
	where $\gamma$ is the systemic velocity, or the radial velocity of the center of mass of the binary system.
	It relates to the motion of the whole system.
	The $\gamma$ always equals 0 in our method since the radial velocities measured by template-matching approach within a source are relative radial velocities, which were not include the systemic velocity.
	$a$ is the long axis of the elliptic orbit of the primary, which is associated with mass ratio $q$.
	$\theta$ is the position angle between the radius vector and a reference direction, which as a function of the time of an observational epoch.
	$P$, $e$, $i$, $\omega$ and $T_{0}$ are defined in Table \ref{table:orbital parameters}. 
	We thus can get a simulated RV by given a set of the 7 orbital parameters and an observational time.
	
	We implemented the algorithm described by S13.
	The criteria is based on the detectable obvious change in RVs.
	A source is deemed a spectroscopic binary star when there is at least one pair of RV satisfies 
	\begin{equation}  \label{Eq:binary_distinguish}
		\frac{\left|v_{i}-v_{j}\right|}{\sqrt{\varepsilon_{i}^{2}+\varepsilon_{j}^{2}}}>4.0 \quad \text { and } \quad\left|v_{i}-v_{j}\right|>C
		,
	\end{equation}
	where $v_{i}$ and $\varepsilon_{i}$ are the 
	RV and its error of epoch $i$ for a given object.
	The value of $C$ was conservatively adopted with $20$ km\,s$^{-1}$ because not only photospheric variations in supergiants can mimic variations with amplitudes of up to $20$ km\,s$^{-1}$ (\citealt{2009A&A...507.1585R}) but also wind effects of some stars (\citealt{2016Natur.529..502L}, \citealt{2020Natur.580E..11A}).
	Even though, it is somewhat subjective to give an exact value to $C$ as a cutoff for binary detection. But, the intrinsic binary fraction $f_{bin}$ of real observations will be corrected through the simulation procedure described as below.
	Therefore, a reasonable selection of $C$ will not significantly affect the final result.
	For a source with more than 3 observations, several pairs of RVs may satisfy Eq.  \ref{Eq:binary_distinguish}, each pairs corresponds to its own observational time scale $\Delta HJD$.
	The maximum RV variation $\Delta RV$ and the minimum time scale $\Delta HJD$ were recorded respectively in two sequences, which will be used to 
	compare with the simulations utilizing
	Kolmogorov-Smirnov (KS) tests.
	
	Accounting for observational biases due to sampling and measurement uncertainties we need to  search for sets of distributions that reproduce the properties of the observations in three aspects: 1) The observed binary fraction; 2) the peak-to-peak amplitude $\Delta RV$ of the RV variations; 3) the minimum time scale $\Delta HJD$ for significant RV variation to be observed.
	Simulated and observed distributions are compared by mean of KS tests.
	The binary fraction detected in the simulations will be compared to the observed fraction using a Binomial distribution \cite{2012Sci...337..444S}. 
    A global merit function ($\Xi^{\prime}$) is constructed as:
	\begin{equation} \label{Eq:global merit function_Sana}
		\Xi^{\prime}=P_{\mathrm{KS}}(\Delta RV) \times P_{\mathrm{KS}}(\Delta HJD) \times B\left(N_{bin}, N, f_{bin}^{simul}\right) ,
	\end{equation}
	where $f_{bin}^{simul}$ and $N_{bin}$ are the binary fraction and the number of samples detected in simulated samples.
	N is the sample size.
	Following S13, we use
	power-law functions
	to describe the intrinsic distributions of orbital parameters, e.g., $f\left(\log_{10}P/\mathrm{d}\right) \sim\left(\log_{10}P\right)^{\pi}$, $f(q) \sim q^{\kappa}$ and $f(e) \sim e^{\eta}$ ($\eta$ is fixed at $-0.5$).
	We thus can explore the distribution of $\Xi^{\prime}$ in the three-dimensional parameter space $(\pi, \kappa, f_{bin})$ by using Monte-Carlo approach.
	
	\subsection{Validation}	\label{Validation}
	\subsubsection{Verification of the algorithm with S13 data}	\label{subsubsec:Experiments with mock data from Sana}
\begin{figure}
		\centering
		\includegraphics[width=1\linewidth]{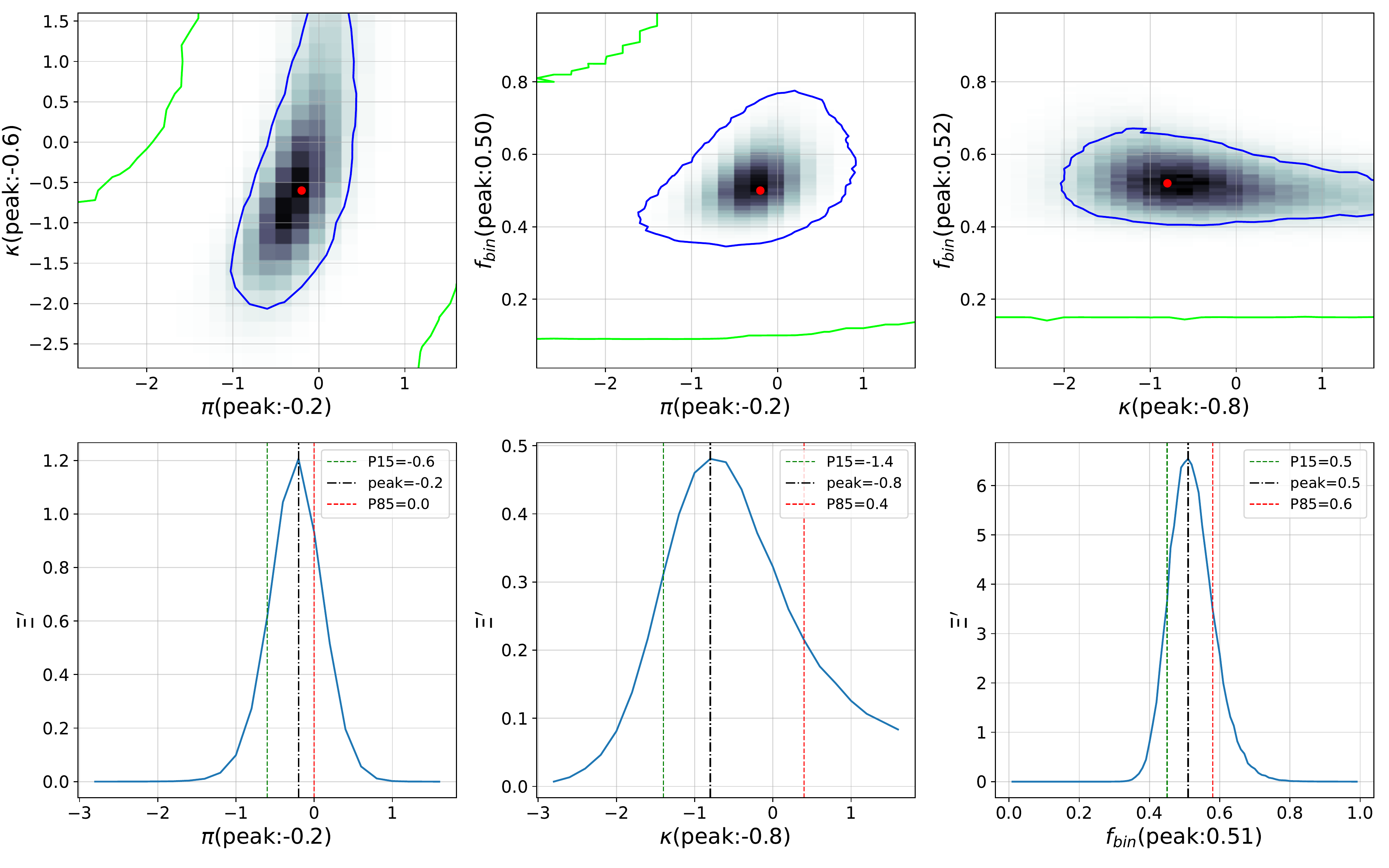}
		\caption{
		The figure shows the results using S13 data. 
		Panels in the top row display the projection of $\Xi^{\prime}$ onto $\pi$ vs. $\kappa$, $\pi$ vs. $f_{bin}$ and $\kappa$ vs. $f_{bin}$ respectively. 
		The green and blue contour lines indicate the 15 and 85 percentile of $\Xi^{\prime}$.
		Panels in the bottom row show the projection of $\Xi^{\prime}$ onto $\pi$, $\kappa$ and $f_{bin}$ respectively.
		The green and red dotted lines indicate the position of 15 and 85 percentile of $\Xi^{\prime}$.
		}
		\label{fig:pi_kappa_fbin_sana_fig4.pdf}
	\end{figure}

	We used the real observational data from S13 to test our code. 
	Following S13, we adopted the detected range of $\log_{10}P$, $q$ and $e$ are $\log_{10}P\in[0.15,3.5]$, $q\in[0.1,1.0]$ and $e\in[10^{-5},0.9]$, respectively.
	The results of our program are shown in Fig.\ref{fig:pi_kappa_fbin_sana_fig4.pdf}.
	Panels in the top row are the projections of $\Xi^{\prime}$ from the multi-dimensional space onto  $\pi$ vs. $\kappa$, $\pi$ vs. $f_{bin}$, and $\kappa$ vs. $f_{bin}$.
	Red points indicate the peak positions.
	In Fig.\ref{fig:pi_kappa_fbin_sana_fig4.pdf}, the bottom panels are one-demensional $\Xi^{\prime}$ projections to $\pi$, $\kappa$, and $f_{bin}$, respectively.

	It is worthy to note that assuming the binary mass ratio follow a power-law distribution may not be very precise (\citealt{2019MNRAS.490..550L}).
	However, it could tell us approximately about the relative amount of binaries with low and high mass ratio to some extent.
	
	The best-fit parameters displayed in the figure well agree with results by S13.
	That means we successfully implemented the algorithm proposed by S13.
	
	\subsubsection{Effect of the true $f_{bin}$}	\label{subsec:Effect of the true fbin}
	More simulations were carried out to investigate how binary fraction affects the results.
	We generated 332 mock samples of binaries with $\pi, \kappa, \eta = (-0.45, -1, -0.5)$ (S13) and 332 samples of single stars with measurement error of each spectrum.
	The number of observations, observational time interval and RV measurement error of simulated binary samples are came from S13 data.
	The primary mass, $m_1$, was randomly drawn from an initial mass function (IMF) described by \cite{1955ApJ...121..161S} with mass range from 15 to 80 $M_{\odot}$.
	The rest parameters $i, \omega, T_{0}$ are randomly drawn from the assumed distributions:
	\begin{equation} \label{Eq:distributions of orbital parameters}
		\cos i \sim Uniform(0,1), \quad
		\omega \sim Uniform(0,2\pi), \quad
		T_{0}\sim Uniform(0,P). 
	\end{equation}
	
	In order to investigate whether the adopted method can accurately estimate  $f_{bin}$, which is the most important parameter, we run the Monte-Carlo procedure with a parameter grid where $\pi\in[-3,2]$, $\kappa\in[-3,2]$. And all the following calculations we adopted the fixed $\eta=-0.5$.
	\begin{figure}
		\centering
		\includegraphics[width=1\linewidth]{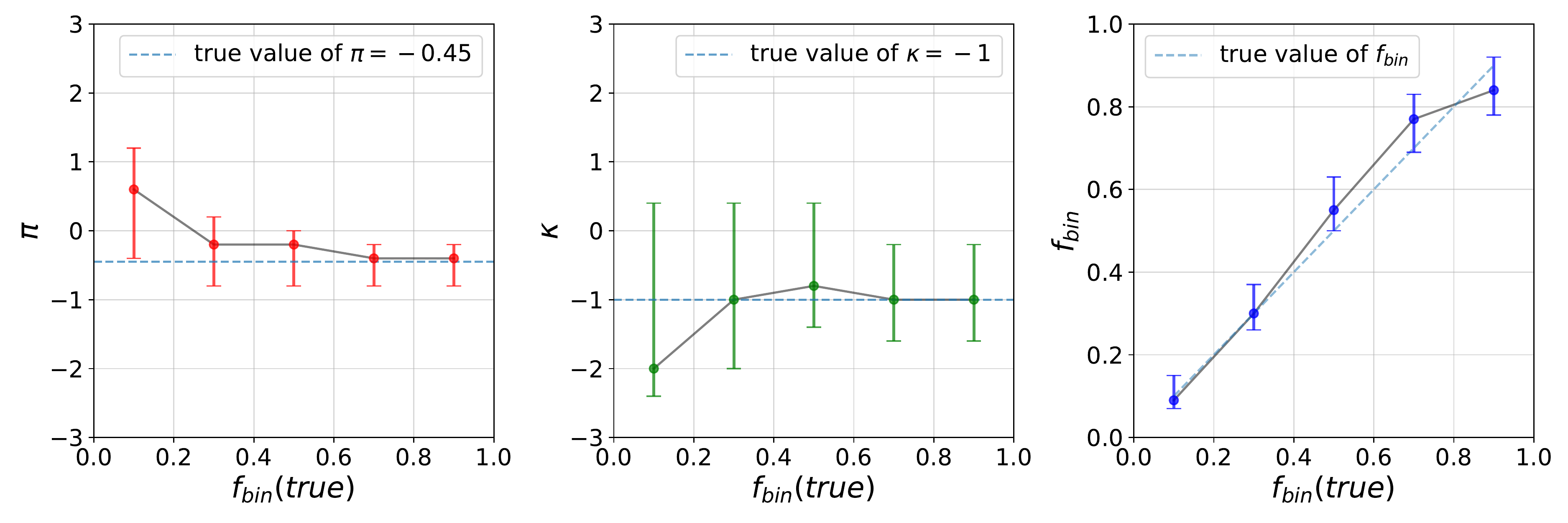}
		\caption{Results of simulation experiments with mock samples generated based on S13 data. 
		The three panels display the recoveries of $\pi$, $\kappa$ and $f_{bin}$ respectively with different true values of $f_{bin}$, short transverse lines at the bottom and top of each vertical line indicated the position of 15 and 85 percentile, points between them are peak positions.
		}
		\label{fig:testComparisonSana_fig5.pdf}
	\end{figure}
	The simulated samples 
	are given with different initial $f_{bin}$ ranging from $0.1$ to $0.9$ with a step of $0.2$.
	The results of $f_{bin}$ estimation and predictions of $\pi, \kappa$ are shown in Fig. \ref{fig:testComparisonSana_fig5.pdf}. The error bars in each panel indicate the $15$ and $85$ percentiles, points between top and bottom of error bars are the peak position.
	
	Fig.\ref{fig:testComparisonSana_fig5.pdf} shows that when $f_{bin}$ is low, the predictions of $\pi$ and $\kappa$ are inaccurate.
	The errors of $\pi$ and $\kappa$ decrease and the predicted values of $\pi$ and $\kappa$ tend to be pinned down at around the ground true values with the increase of $f_{bin}$.
	This is in all probability because the estimations of orbital distributions rely on the enough fraction of binary systems.
	
	Our exercise shows that when $f_{bin}>0.3$ the estimation of $\pi$ and $\kappa$ is reliable. 
	From the right panel we can see the the $f_{bin}$ prediction is more accurate than $\pi$ and $\kappa$.
	This demonstrates that the method is credible at least in the $f_{bin}$ estimation for the mock samples.
	
	\subsubsection{Validation with mock data mimic LAMOST samples}	\label{subsec:Validation with mock data mimic LAMOST samples}
\begin{figure}
		\centering
		\includegraphics[width=1\linewidth]{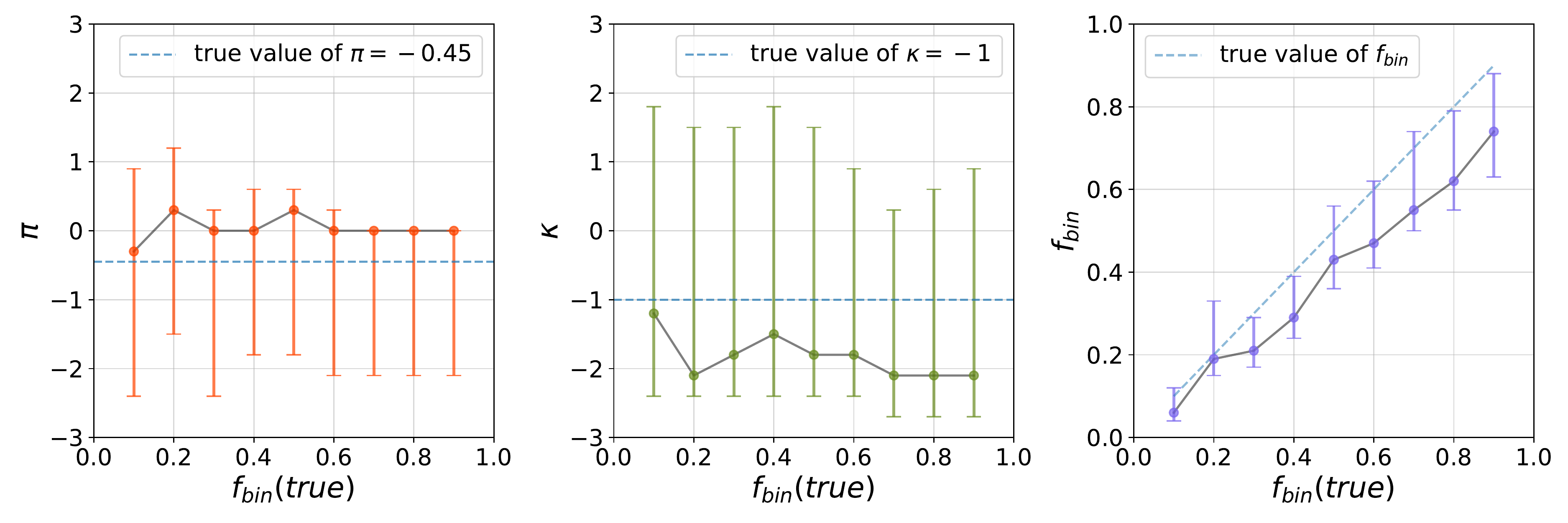}
		\caption{
		Results of simulations with mock samples generated based on LAMOST OB samples.
		The three panels display the recoveries of $\pi$, $\kappa$ and $f_{bin}$ respectively with different true values of $f_{bin}$, short transverse lines at the bottom and top of each vertical line indicated the position of 15 and 85 percentile, points between them are peak positions.
		}
		\label{fig:testComparisonLamost_fig6.pdf}
	\end{figure}

	We then did the third simulations to investigate whether the LAMOST samples fit the method.
	Mock samples mimic LAMOST OB stars with different $f_{bin}$ from $0.1$ to $0.9$ with step of $0.1$, were generated in the same way mentioned above but the range of $m_1$ was from 5 to 20 $M_{\odot}$. 
	The differences are the value range of $\log_{10}P$ is $[0.15,3.0]$, and the detection range of $\pi, \kappa$ are $\pi\in[-3,3], \kappa\in[-3,3]$.
	The corresponding results are shown in Fig. \ref{fig:testComparisonLamost_fig6.pdf}.
	
	The left panel displays the estimation of $\pi$, from which we found that there are evident asymmetry and slight overestimates in almost all samples with different $f_{bin}$.
	When $f_{bin} = 0.7, 0.8, 0.9$, the corresponding peak values of $\pi$ overlapped with their correspond 85 percentiles meaning that the probability distribution of $\pi$ is highly skewed toward large values.
	The estimation of $\kappa$, delineated in the middle panel, shows that there also are obvious asymmetry and underestimations in all cases.
	Compared with S13 results we find the peak value of $\kappa$ is skewed toward smaller values.
	But the skewness may be also caused by the limit of detection range we defined.
	
	From the right panel we can see that $f_{bin}$ estimates are roughly around the ground trueth in all cases. 
	However, the derived $f_{bin}$ are slightly underestimated by around 0.1 when the real $f_{bin}$ is smaller than 0.6.
	When the true value is larger than 0.6, the estimates are underestimated by about 0.15.
	The different behavior of $f_{bin}$ between Fig.\ref{fig:testComparisonSana_fig5.pdf} and Fig.\ref{fig:testComparisonLamost_fig6.pdf} is likely due to the difference of the two dataset especially their differences in observation epochs.
	\section{Result} \label{sec:Results}
	\subsection{Estimation of $f_{bin}$ with open $\pi$ and $\kappa$}	\label{subsec:Estimation of fbin with open pi and kappa}
	
	\begin{figure}
		\centering
		\includegraphics[width=1\linewidth]{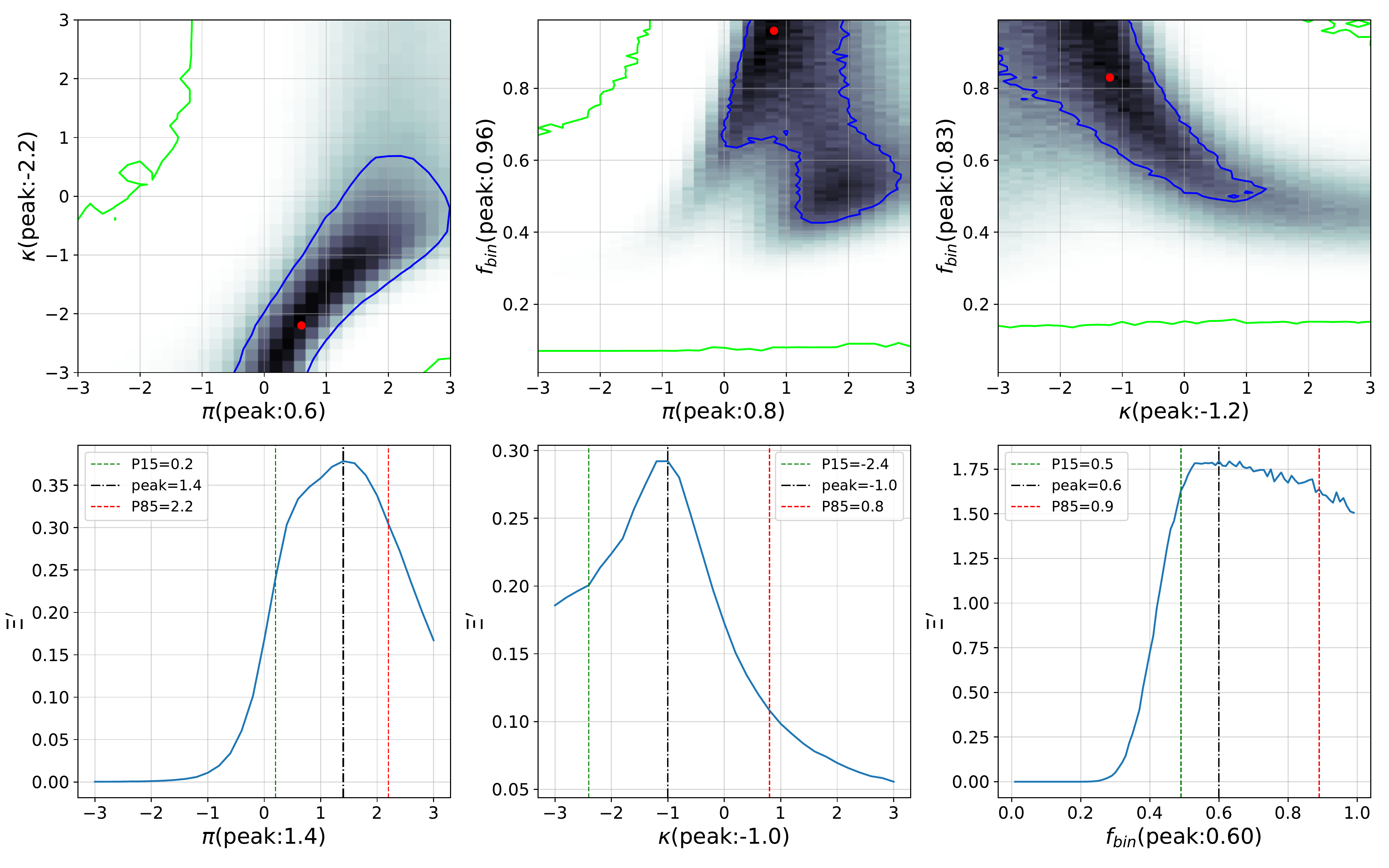}
		\caption{Results of the LAMOST 328 OB stars with at least 3 observations. 
		Panels in the top row display the projection of $\Xi^{\prime}$ onto $\pi$ vs. $\kappa$, $\pi$ vs. $f_{bin}$ and $\kappa$ vs. $f_{bin}$ respectively. 
		The green and blue contour lines indicate the 15 and 85 percentile of $\Xi^{\prime}$.
		Panels in the bottom row show the projection of $\Xi^{\prime}$ onto $\pi$, $\kappa$ and $f_{bin}$ respectively.
		The green and red dotted lines indicate the position of 15 and 85 percentile of $\Xi^{\prime}$.
		}
		\label{fig:pi_kappa_fbin_lamost_fig7.pdf}
	\end{figure}
	By mean of Eq.\ref{Eq:binary_distinguish}, 72 samples were satisfied and are regarded as binary in the 328 OB samples selected in Sect. \ref{subsec:Samples from LAMOST}.
	So the observed binary fraction is $72/328=0.22$.
	The calculated results are shown in Fig.\ref{fig:pi_kappa_fbin_lamost_fig7.pdf}, which layout is similar with Fig.\ref{fig:pi_kappa_fbin_sana_fig4.pdf}.
	
	As shown in the top-left panel, the $\Xi^{\prime}$ range larger than 85 percentile is larger compared to S13 result.
	The distribution of $\Xi^{\prime}$ in $\pi$ vs. $\kappa$ indicates that the two parameters are correlated to some extent.
	It seems that $\Xi^{\prime}$ shows two peaks with irregular shape in top-middle panel.
	
	The $f_{bin}$ and $\kappa$ are anti-correlated especially when $\kappa<0$ as shown in top-right panel.
	A lower $\kappa$ means that our model considers more binaries with a lower $q$ than those have a higher $q$.
	When the model chooses smaller $\kappa$ more binaries having much lower mass companion.
	In this case, single stars are easy to contaminate to binary samples.
	As a consequence, $f_{bin}$ is likely larger.
	This can well explain the anti-correlation shown in $\kappa$ vs. $f_{bin}$ plane.
	The bottom panels display the marginal $\Xi^{\prime}$ of $\pi, \kappa$ and $f_{bin}$.
	All the 3 distributions are much broader compared with S13 (also Fig.\ref{fig:pi_kappa_fbin_sana_fig4.pdf}).
	The peak positions of $\pi$ and $\kappa$ are $1.4$ and $-1$ (bottom-left and bottom-middle panels in Fig.\ref{fig:pi_kappa_fbin_lamost_fig7.pdf}).
	A larger peak value of $\pi$ means there are relative more binaries with long orbital periods than S13 samples.
	This may be also caused by fewer observations so that gives a partial result with little statistical significance.
	The estimation of $\kappa$ is essentially consistent with S13 with a larger uncertainty.
	
	It seems that the current samples, with less numbers of observations, are hard to simultaneously constrain $\pi$, $\kappa$ and $f_{bin}$.
	We then turn to only constrain $f_{bin}$ with the Monte-Carlo procedure by fixing $\pi$ and $\kappa$ at $-0.45$ and $-1$, respectively, as concluded by S13.

	\subsection{Estimation of $f_{bin}$ with fixed $\pi$ and $\kappa$}	\label{subsec:Estimation of fbin with fixed pi and kappa}
	
	\begin{figure}[htbp]
		\centering
		\includegraphics[width=0.85\linewidth]{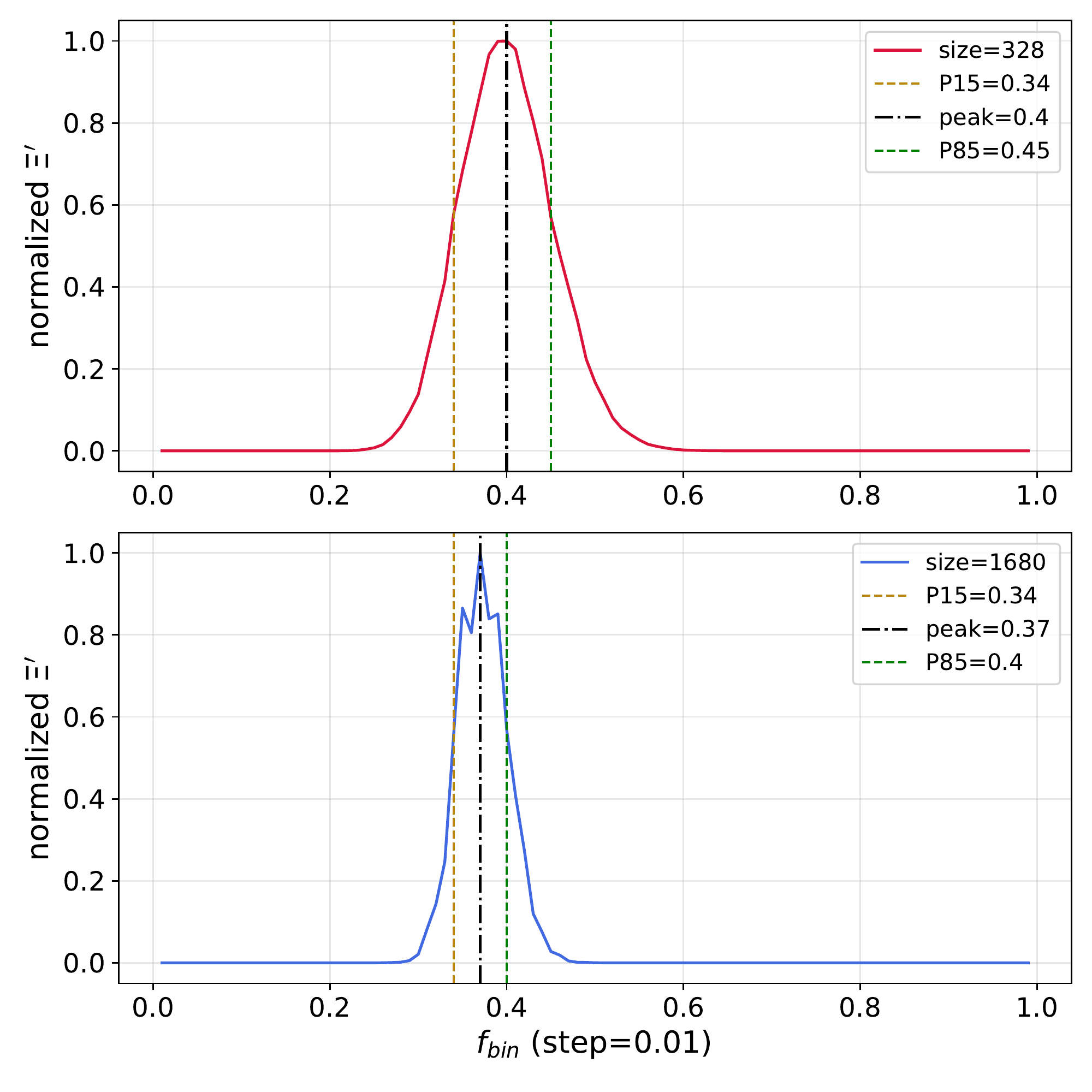}
		\caption{Estimation of $f_{bin}$ under the condition $(\pi,\kappa)=(-0.45,-1)$. The top panel: It displays the result of subset containing 328 samples. The 15 and 85 percentiles are 0.34 and 0.45 respectively, and the peak position at 0.4. The bottom panel: It shows the result of whole samples containing 1680 stars. The 15 and 85 percentiles are 0.34 and 0.4 respectively, and the peak position at 0.37.}
		\label{fig:likelihood_fbin_fixedpar_mc200000_fig8.pdf}
	\end{figure}
	
	If $\pi$ and $\kappa$ are fixed, a few experiments we made show that the fluctuation of statistics is remarkable when the number of random samplings in Monte-Carlo procedure is 100.
	In order to weaken the statistical fluctuation, the steps of Monte-Carlo procedure was increased to $2\times10^{5}$.
	According to Eq. \ref{Eq:RV}, a multi-degree-of-freedom formula implies that more repeated observations were need to predict the distributions of orbital periods and mass ratio.
	But, the $f_{bin}$ could be estimated when the parameters, $\pi$ and $\kappa$, are fixed.
	The samples excluded in Sect.\ref{sec:Data} with 2 observations can be used under this condition.
	So we calculated the total 1680 samples at the same time.
	The observed binary fraction directly derived by Eq. \ref{Eq:binary_distinguish} of total samples is 0.15.
	Calculated results of subset and total samples were shown in Fig. \ref{fig:likelihood_fbin_fixedpar_mc200000_fig8.pdf}.
	Where the ordinate indicated the normalized $\Xi^{\prime}$, and $f_{bin}$ is $0.4_{-0.06}^{+0.05}$ for the subset, and $0.37_{-0.03}^{+0.03}$ for the total samples.
	The error bar become narrower when number of samples goes up, and the closer peaks shown that the two results are consistent with each other.
    
    Note that, with LAMOST data the adopted S13 method tends to underestimate $f_{bin}$ by about 0.15 (see Fig.\ref{fig:testComparisonLamost_fig6.pdf}).
    Therefore, the real $f_{bin}$ of the OB samples may reach $\sim$ 0.65.
    However, since the simulations to produce Fig.\ref{fig:testComparisonLamost_fig6.pdf} are under different conditions, precise systematic correction for $f_{bin}$ is difficult.
    
	\section{Discussion} \label{sec:Discussion}
	\subsection{Number of observation} \label{sec:Number of observation}
	\begin{figure}[htbp]
		\centering
		\includegraphics[width=1\linewidth]{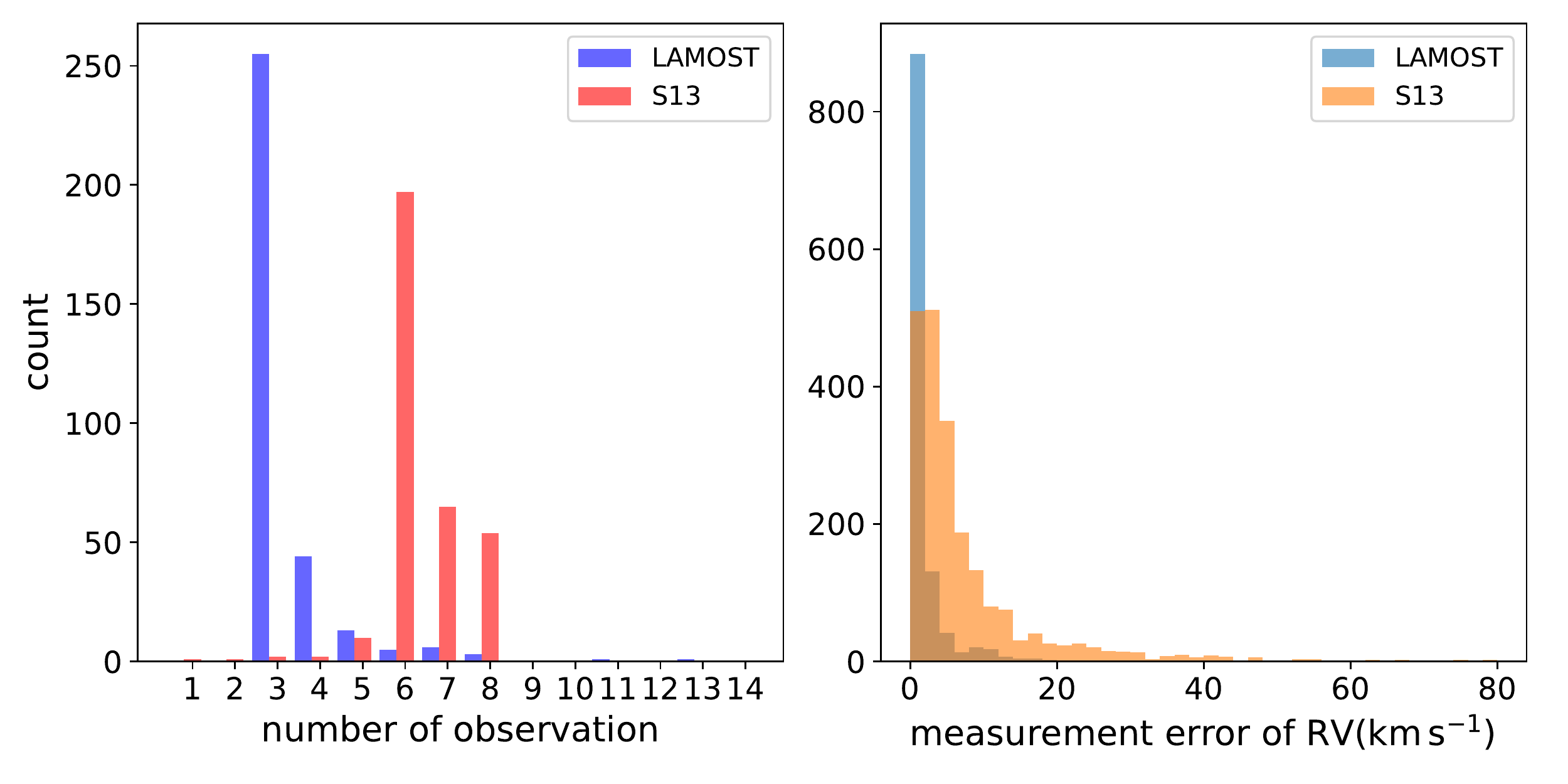}
		\caption{Left panel: It displays the comparison of number of observation between data from S13 and LAMOST, blue bars indicated LAMOST data, and red bars represented S13 data.
			Right panel:It depicts the comparison of measurement error of RV between data from S13 and LAMOST, cyan bars indicated LAMOST data, and orange bars represented S13 data.}
		\label{fig:obstmsMmeComparison_fig9.pdf}
	\end{figure}
	In the top panels of Fig.\ref{fig:pi_kappa_fbin_lamost_fig7.pdf}, it is seen that constraints on $\pi$,$\kappa$ and $f_{bin}$ are not very precise compared with results of S13 shown in Fig.\ref{fig:pi_kappa_fbin_sana_fig4.pdf}. 
	That's because there are 7 parameters, ($P,q,e,m_{1},i,\omega,T_{0}$), and the time 
    of an observational epoch, were needed to derive a RV from Eq.\ref{Eq:RV}.
	Theoretically, at least 7 observations for a given source are indispensable to provide sufficient information helping to the parameters determination.
	However, there are $255$ sources in our $328$ samples with exactly 3 observations.
	Fewer information they carried leads to larger uncertainty of the orbital parameter estimates.
	Number of observations of S13 and our samples are displayed in the left panel in  Fig.\ref{fig:obstmsMmeComparison_fig9.pdf}.
	Most of S13 samples have $6,7$ and $8$ repeated observations, but $3,4$ and $5$ in our samples.
	That's the main reason our results is limited by larger uncertainty.
	
	\subsection{Measurement error of relative RV} \label{sec:Measurement error of relative RV}
	While selecting sources from LAMOST, we filtered samples whose RV measurement error lager than $20$ km\,s$^{-1}$.
	The abandoned samples generally have spectra with low SNRg or defective pixels.
	The samples with larger RV measurements error were contained in S13 data and thus their RV variation may be not very accurate, see right panel of  Fig.\ref{fig:obstmsMmeComparison_fig9.pdf}.
	
	\subsection{Cutoff of orbital period} \label{sec:Cutoff of orbital period}
	The maximum of detection range of $\log_{10}P$, or $P$, obviously affect the final estimation of $f_{bin}$.
	A smaller maximum of $P$ means the model can not detect binaries with orbital periods longer than the maximum.
	Then, these binaries will be regarded as single stars, and the $f_{bin}$ goes smaller at the same time.
	
	The cutoff of detection range of $\log_{10}P$ of our samples is 3.0, which is smaller than 3.5 adopted by S13.
	So the final calculated $f_{bin}$ results is smaller than S13 naturally.
	But, the cutoff of $P$ is not the larger the better. 
	Due to the influences from measurement error of RV, single stars have more risk to fall into binaries when the detection range become wider.  
	
	\section{Conclusion} \label{sec:conclusion}
	We applied the method introduced by S13 to estimate the properties of binary stars based on the LAMOST OB samples.
	We find $f_{bin}$ is $0.4_{-0.06}^{+0.05}$ for the samples with at least 3 observations and $0.37_{-0.03}^{+0.03}$ for the total samples.
	Many simulations were made to investigate the capability of the S13 method.
	We find that because the epochs of the LAMOST observations are smaller than S13 samples, $\pi$ and $\kappa$ can not be well constrained.
	In future, we will revisit this problem with LAMOST time-domain data so that $\pi$ and $\kappa$ can also be estimated.
	
	
	
	\begin{acknowledgements}
	    This work is supported by the National Key R\&D Program of China No. 2019 YFA0405501.
	    This work is also supported by the National Natural Science Foundation of China with grant Nos. 11835057, 12090040, 12090043.
		Guoshoujing Telescope (the Large Sky Area Multi-Object Fiber Spectroscopic Telescope LAMOST) is a National Major Scientific Project built by the Chinese Academy of Sciences. 
		Funding for the project has been provided by the National Development and Reform Commission. LAMOST is operated and managed by the National Astronomical Observatories, Chinese Academy of Sciences. 
		This research uses data obtained through LAMOST, which is operated and managed by the National Astronomical Observatories, Chinese Academy of Sciences, and the Special Fund for Astronomy from the Ministry of Finance.
		\textit{Facilities}: LAMOST
		\textit{Software}: astropy(\citealt{2018AJ....156..123A}),scipy(\citealt{2020NatMe..17..261V}),TOPCAT(\citealt{2005ASPC..347...29T})
		
	\end{acknowledgements}

	\bibliographystyle{raa}
	\bibliography{bibtex}

\begin{thebibliography}{33}
\providecommand\natexlab[1]{#1}
\providecommand\JournalTitle[1]{#1}

\bibitem[{Abdul-Masih} {et~al.}(2020)]{2020Natur.580E..11A}
{Abdul-Masih}, M., {Banyard}, G., {Bodensteiner}, J., {et~al.} 2020, \nat, 580,
  E11

\bibitem[{Aldoretta} {et~al.}(2015)]{2015AJ....149...26A}
{Aldoretta}, E.~J., {Caballero-Nieves}, S.~M., {Gies}, D.~R., {et~al.} 2015,
  \aj, 149, 26

\bibitem[{Almeida} {et~al.}(2017)]{2017A&A...598A..84A}
{Almeida}, L.~A., {Sana}, H., {Taylor}, W., {et~al.} 2017, \aap, 598, A84

\bibitem[{Astropy Collaboration} {et~al.}(2018)]{2018AJ....156..123A}
{Astropy Collaboration}, {Price-Whelan}, A.~M., {Sip{\H{o}}cz}, B.~M., {et~al.}
  2018, \aj, 156, 123

\bibitem[{Badenes} {et~al.}(2018)]{2018ApJ...854..147B}
{Badenes}, C., {Mazzola}, C., {Thompson}, T.~A., {et~al.} 2018, \apj, 854, 147

\bibitem[{Chini} {et~al.}(2012)]{2012MNRAS.424.1925C}
{Chini}, R., {Hoffmeister}, V.~H., {Nasseri}, A., {Stahl}, O., \& {Zinnecker},
  H. 2012, \mnras, 424, 1925

\bibitem[{Cui} {et~al.}(2012)]{2012RAA....12.1197C}
{Cui}, X.-Q., {Zhao}, Y.-H., {Chu}, Y.-Q., {et~al.} 2012, Research in Astronomy
  and Astrophysics, 12, 1197

\bibitem[{Dunstall} {et~al.}(2015)]{2015AA...580A..93D}
{Dunstall}, P.~R., {Dufton}, P.~L., {Sana}, H., {et~al.} 2015, \aap, 580, A93

\bibitem[{Eldridge} {et~al.}(2011)]{2011MNRAS.414.3501E}
{Eldridge}, J.~J., {Langer}, N., \& {Tout}, C.~A. 2011, \mnras, 414, 3501

\bibitem[{Fernandez} {et~al.}(2017)]{2017PASP..129h4201F}
{Fernandez}, M.~A., {Covey}, K.~R., {De Lee}, N., {et~al.} 2017, \pasp, 129,
  084201

\bibitem[{Gao} {et~al.}(2017)]{2017MNRAS.469L..68G}
{Gao}, S., {Zhao}, H., {Yang}, H., \& {Gao}, R. 2017, \mnras, 469, L68

\bibitem[{Kobulnicky} {et~al.}(2014)]{2014ApJS..213...34K}
{Kobulnicky}, H.~A., {Kiminki}, D.~C., {Lundquist}, M.~J., {et~al.} 2014,
  \apjs, 213, 34

\bibitem[{Langer} {et~al.}(2008)]{2008IAUS..250..167L}
{Langer}, N., {Cantiello}, M., {Yoon}, S.-C., {et~al.} 2008, in Massive Stars
  as Cosmic Engines, ed. F.~{Bresolin}, P.~A. {Crowther}, \& J.~{Puls}, Vol.
  250, 167

\bibitem[{Li} {et~al.}(2016)]{2016Natur.529..502L}
{Li}, C., {de Grijs}, R., {Deng}, L., {et~al.} 2016, \nat, 529, 502

\bibitem[{Liu}(2019)]{2019MNRAS.490..550L}
{Liu}, C. 2019, \mnras, 490, 550

\bibitem[{Liu} {et~al.}(2019)]{2019ApJS..241...32L}
{Liu}, Z., {Cui}, W., {Liu}, C., {et~al.} 2019, \apjs, 241, 32

\bibitem[{Luo} {et~al.}(2015)]{2015RAA....15.1095L}
{Luo}, A.~L., {Zhao}, Y.-H., {Zhao}, G., {et~al.} 2015, Research in Astronomy
  and Astrophysics, 15, 1095

\bibitem[{Merle} {et~al.}(2017)]{2017A&A...608A..95M}
{Merle}, T., {Van Eck}, S., {Jorissen}, A., {et~al.} 2017, \aap, 608, A95

\bibitem[{Minor}(2013)]{2013ApJ...779..116M}
{Minor}, Q.~E. 2013, \apj, 779, 116

\bibitem[{Moe} \& {Di Stefano}(2017)]{2017ApJS..230...15M}
{Moe}, M., \& {Di Stefano}, R. 2017, \apjs, 230, 15

\bibitem[{Podsiadlowski} {et~al.}(1992)]{1992ApJ...391..246P}
{Podsiadlowski}, P., {Joss}, P.~C., \& {Hsu}, J.~J.~L. 1992, \apj, 391, 246

\bibitem[{Pourbaix} {et~al.}(2004)]{2004A&A...424..727P}
{Pourbaix}, D., {Tokovinin}, A.~A., {Batten}, A.~H., {et~al.} 2004, \aap, 424,
  727

\bibitem[{Price-Whelan} {et~al.}(2017)]{2017ApJ...837...20P}
{Price-Whelan}, A.~M., {Hogg}, D.~W., {Foreman-Mackey}, D., \& {Rix}, H.-W.
  2017, \apj, 837, 20

\bibitem[{Raghavan} {et~al.}(2010)]{2010ApJS..190....1R}
{Raghavan}, D., {McAlister}, H.~A., {Henry}, T.~J., {et~al.} 2010, \apjs, 190,
  1

\bibitem[{Ritchie} {et~al.}(2009)]{2009A&A...507.1585R}
{Ritchie}, B.~W., {Clark}, J.~S., {Negueruela}, I., \& {Crowther}, P.~A. 2009,
  \aap, 507, 1585

\bibitem[{Salpeter}(1955)]{1955ApJ...121..161S}
{Salpeter}, E.~E. 1955, \apj, 121, 161

\bibitem[{Sana} {et~al.}(2012)]{2012Sci...337..444S}
{Sana}, H., {de Mink}, S.~E., {de Koter}, A., {et~al.} 2012, Science, 337, 444

\bibitem[{Sana} {et~al.}(2013)]{2013A&A...550A.107S}
{Sana}, H., {de Koter}, A., {de Mink}, S.~E., {et~al.} 2013, \aap, 550, A107

\bibitem[{Sota} {et~al.}(2014)]{2014ApJS..211...10S}
{Sota}, A., {Ma{\'\i}z Apell{\'a}niz}, J., {Morrell}, N.~I., {et~al.} 2014,
  \apjs, 211, 10

\bibitem[{Taylor}(2005)]{2005ASPC..347...29T}
{Taylor}, M.~B. 2005, in Astronomical Society of the Pacific Conference Series,
  Vol. 347, Astronomical Data Analysis Software and Systems XIV, ed.
  P.~{Shopbell}, M.~{Britton}, \& R.~{Ebert}, 29

\bibitem[{Troup} {et~al.}(2016)]{2016AJ....151...85T}
{Troup}, N.~W., {Nidever}, D.~L., {De Lee}, N., {et~al.} 2016, \aj, 151, 85

\bibitem[{Virtanen} {et~al.}(2020)]{2020NatMe..17..261V}
{Virtanen}, P., {Gommers}, R., {Oliphant}, T.~E., {et~al.} 2020, Nature
  Methods, 17, 261

\bibitem[{Zhao} {et~al.}(2012)]{2012RAA....12..723Z}
{Zhao}, G., {Zhao}, Y.-H., {Chu}, Y.-Q., {Jing}, Y.-P., \& {Deng}, L.-C. 2012,
  Research in Astronomy and Astrophysics, 12, 723

\end{thebibliography}
	
\end{document}